\def\year{2017}\relax
\documentclass[letterpaper]{article} 
\usepackage{aaai17}  
\usepackage{times}  
\usepackage{helvet}  
\usepackage{courier}  
\usepackage{url}  
\usepackage{graphicx}  
\begin{document}
%
\title{Detecting the Hate Code on Social Media}
\author{Rijul Magu \\University of Rochester \\rmagu2@cs.rochester.edu
\And
Kshitij Joshi \\University of Rochester \\kjoshi4@cs.rochester.edu
\And
Jiebo Luo \\University of Rochester \\jluo@cs.rochester.edu}
\maketitle
\begin{abstract}
Social media has become an indispensable part of the everyday lives of millions of people around the world. It provides a platform for expressing opinions and beliefs, communicated to a massive audience. However, this ease with which people can express themselves has also allowed for the large scale spread of propaganda and hate speech. To prevent violating the abuse policies of social media platforms and also to avoid detection by automatic systems like Google's Conversation AI, racists have begun to use a “code” (a movement termed “Operation Google”). This involves substituting references to communities by benign words that seem out of context, in hate filled posts or Tweets. For example, users have used the words “Googles” and ​“Bings” to represent the African-American and Asian communities, respectively.  By generating the list of users who post such content, we move a step forward from classifying tweets by allowing us to study the usage pattern of these concentrated set of users.
\end{abstract}

\section{Introduction}
Internet usage has grown tremendously over time. From a figure of 52\% of the American population that used the internet in 2000, the percentage of users grew to 84\% in a span of 15 years \cite{perrin2015americans}. Within that time frame, we witnessed the rise of social media: a path-breaking socio-technological concept that changed the way people exchanged ideas and broadcast opinions. Social media  has seen a parallel jump in popularity, with over 65\% of American adults using social media websites in 2015, as opposed to 7\% in 2005 \cite{perrin2015social}. While this increase in the user base has produced a world that is more connected than ever before, it has also had serious negative implications that need to be accounted for.

One of the most prominent ones is the ascent of online hate speech. While hate speech has been around for a long time, the ease of access to reaching and influencing the masses has been augmented through social media. The situation at present is alarming. A 2016 report analyzing antisemitic content on social media exposed the dismal performance of certain social media websites such as Twitter and YouTube with regards to the removal of hate content (Twitter removed only about 22\% antisemitic content over a 10 month period). As Jon Weisman of The New York Times puts it, Twitter in particular has become a “cesspit of hate”.

To compound the issue, a new evasive technique to avoid detection of hate speech has emerged. To circumvent violating the abusive behaviour policies of social media websites such as Twitter, such users have now begun to adopt a code: a list of mappings between the names of intended targeted communities and seemingly innocuous, unrelated terms. The idea is to replace references to these communities in hate posts by their representative code words. Some commonly used code words at present are given in Table 1.

\begin{table}[h!]
\centering
\caption{Some common codewords}
\hfill \break
\begin{tabular}{|c |c|} 
\hline
Code word & Actual word\\ [0.5ex] 
\hline\hline
Google & Black \\
\hline
Yahoo & Mexican \\
\hline
Skype & Jew \\
\hline
Bing & Chinese \\
\hline
Skittle & Muslim \\
\hline
Butterfly & Gay \\
\hline
\end{tabular}

\label{table:1}
\end{table}

\noindent Consider the example,

\begin{center}
\textit{“Gas the skypes”}\\
\end{center}

Here, “skypes” refers to the jewish community in the code. Whereas, in the sentence,

\begin{center}
\textit{“I skype my mom everyday”}\\
\end{center}

The word “skype” likely implies the use of {\it Skype} the Internet service.

In this paper, we first introduce a mechanism to identify instances of such coded hate content. Because of the prominence of this phenomena on Twitter, all of our experimentation is carried out on tweets. We show that we are largely successful in separating the tweets which use the code words in the intended racist context from those that use the words in the regular sense. Furthermore, we create a system that extracts a set of users who frequently use the code to enable us to study their usage patterns. 
 
\section{Related work}
Researchers have in the past studied negative sentiment on social media \cite{bollen2011happiness}. Work has also been done on building classifiers that can identify content that is hateful and antagonistic in nature \cite{burnap2015cyber,silva2016analyzing}. Additionally, techniques have at times involved building a “hate lexicon” to solve the problem. \cite{gitari2015lexicon}. However, most of these projects focus on content that is openly hateful and not disguised to evade detection. Although related, our endeavour and its novelty is geared more towards finding such instances that seek to “fool the system”.

\section{Dataset}
We utilized the Jefferson-Henrique script, a web scraper to collect a total of over a million tweets for a time frame of around a month, starting 23rd September 2016 when reportedly an incident of the first usage of hate-code words was first observed till 18th October 2016, a week after the US Presidential Election, for all the known hate-code words described in Table 1. We ran the script each time with a different code word to pull twitter data specific to that code. Finally, we extracted about a quarter of a million unique English tweets to form the set that we worked with. 

\section{Methodologies for Detecting the Hate\\}
\subsection{\\Preliminary Analysis}
We ran some preliminary experiments to better understand the data before we carried out the main experiment. There were some interesting observations that were noted.

First, by randomly annotating tweets in our database, we extracted 1048 tweets which we could label hateful and 951 tweets which could be labelled non-hateful. Using this set, we calculated the Pearson correlation coefficient between the occurrence of every term present in the set of tweets and the class label. Note that the terms have been lemmatized. The top 10 correlated terms are shown in Table 2.

An analysis of the top 10 terms provide insights into some common themes regarding the hate code. The hashtag '\#MAGA', refers to the slogan 'Make America Great Again' used during the elections. The hashtag '\#MAWA' is a more racist version of this, which stands for 'Make America White Again'- a reference to white nationalism. Additionally, the term 'white' also pops up very high in the list.

A lot of these terms are common phrases and terminology used by the group of people who identify themselves as the alternative right, or the alt-right. The group shares a set of ideologies in the far-right of the American political spectrum, calling for a new, radical conservative movement. The group has been criticized for delving in racism, antisemitism and homophobia, with frequent parallels being drawn with the white supremacist movement. Not surprisingly, therefore, one of the top terms in the list, coming in at second place, is the hashtag '\#ALTRIGHT'.

One alarming observation is the occurrence of the term 'gas'. The term gas is almost always used within hateful tweets as a part of the phrase 'gas the skypes', where 'skypes' refers to Jews. It is particularly this unchecked abject display of hatred and calls for violence that we hoped to capture through our system.

The most interesting of these, however, is the use of the triple parentheses, or the echo symbol. The triple parentheses is used an antisemitic device on Twitter to distinctly mark out the Jewish community or individual Jews, often as a means to bully them. The typical practice is to surround a person's name by the brackets on either side. We noted that significant numbers of people who used the code word 'skype' often also made use of the triple parentheses.

\begin{table}[h!]
\centering
\caption{Top 10 most correlated terms}
\hfill \break
\begin{tabular}{|c|c|} 
\hline
Term & Pearson correlation coefficient \\ [0.5ex] 
\hline\hline
\#MAGA & 0.149 \\
\hline
\#ALTRIGHT & 0.140 \\
\hline
gas & 0.136 \\
\hline
((( ))) & 0.136 \\
\hline
white & 0.136 \\
\hline
war & 0.118 \\
\hline
hate & 0.100 \\
\hline
\#MAWA & 0.098 \\
\hline
destroy & 0.083 \\
\hline
goy & 0.083 \\
\hline
\end{tabular}
\vspace{-15pt}
\label{table:2}
\end{table}

\subsection{\\Identifying Hate Messages}
Next, we ran the Apriori algorithm for frequent itemset mining (Agrawal and Srikant 1994) on a randomly selected set of tweets that we identified as being hateful. Some of the primary frequent patterns that emerged were: [\#MAGA, \#Altright, Skypes], [Gas, Skypes], [Googles, Skittles,Skypes].

Insights worth noting are:
\begin{itemize}
\item The Jewish community is particularly targeted, this can be seen from the fact that many of the frequent itemsets contained the word Skype (a reference to Jews). For example, the phrase “gas the skypes” was a common occurrence.
\item A large number of  tweets contained multiple codewords together. This reflected the fact that aggressors displayed blanket hate towards multiple communities, as opposed to targeted preferences.
\end{itemize}

To this end, the pairwise co-occurrence values as a percentage of the total extracted hateful tweets can be observed in Figure 1. It is noteworthy that the value for [Googles, Skypes] trumps other pairwise co-occurences by a sizable margin, standing at 9.6\% of the total tweets.

\begin{figure}[h]
\centering
\includegraphics[width=\linewidth]{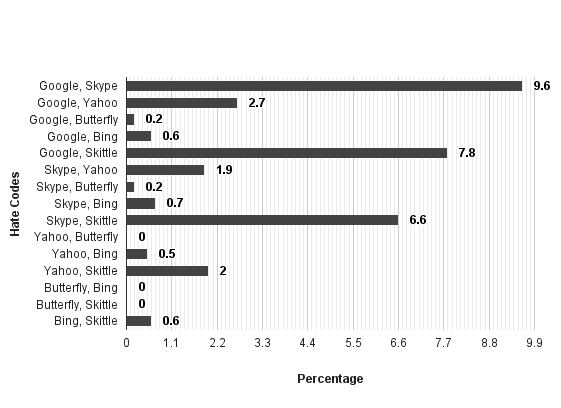}
\caption{Pairwise co-occurrence values in percentage.}
\centering
\end{figure}

From a temporal perspective, we noted a sharp spike in the use of code words in the first week of October, peaking around the 4th of October. This coincided with the second presidential debate. The distribution is shown in Figure 2.

\begin{figure}[h]
\includegraphics[width=\linewidth]{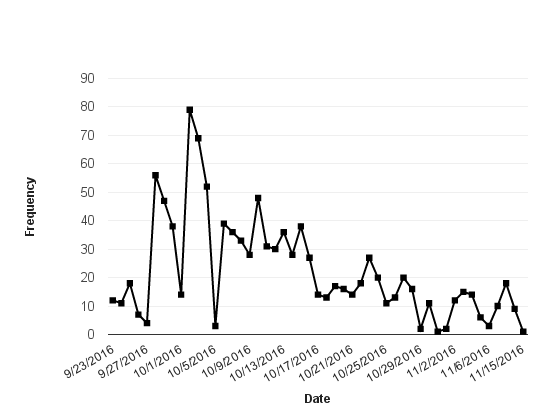}
\caption{Hateful tweets over time.}
\centering
\end{figure}  

\subsection{\\Identifying Aggressors}
An important aspect of our approach consisted of training a classifier to identify hate tweets which included usage of coded words for communities. Our initial step involved annotating a sizable number of tweets to be used as the training set. We randomly picked tweets out of the tweet base and continued annotations till we had a class balance between the class of benign (the negative cases) and hateful tweets (the positive cases). Moreover, we only marked a tweet as hateful if it had a code word for a community and it appeared as if the community was being targeted. All other cases we marked as benign tweets, even those cases in which a code word had been used but not to harshly mock or call for an attack against but to protect or defend a community.

For example:\\
\newline
\noindent \textit{“Imagine a zombie scenario where the zombies are mainly googles \& the scientists who unleash the googles are skypes.”} \\

Clearly communities are being mocked here (it is apparent if one replaces skypes with the jewish people and googles with the Afro-american community and hence it has to be marked as encoded hate speech.

However, for the following examples:\\
\\
\noindent \textit{“Missing my family, but lucky to have one that always skypes me for family dinner.”}

\hfill \break
\noindent \textit{“hey racist piece we know what googles yahoos skypes and skittles are now your account isn't gonna last”} \\

In the first example, skype is indeed used as a benign verb which is meant to indicate the activity of skype-calling. In the second example though, the twitterati evidently did not mean those words to be benign words but communities. But as the person is trying to warn a potential aggressor it was not marked a positive case.\\
 
\begin{figure}[h]
\includegraphics[width=\linewidth]{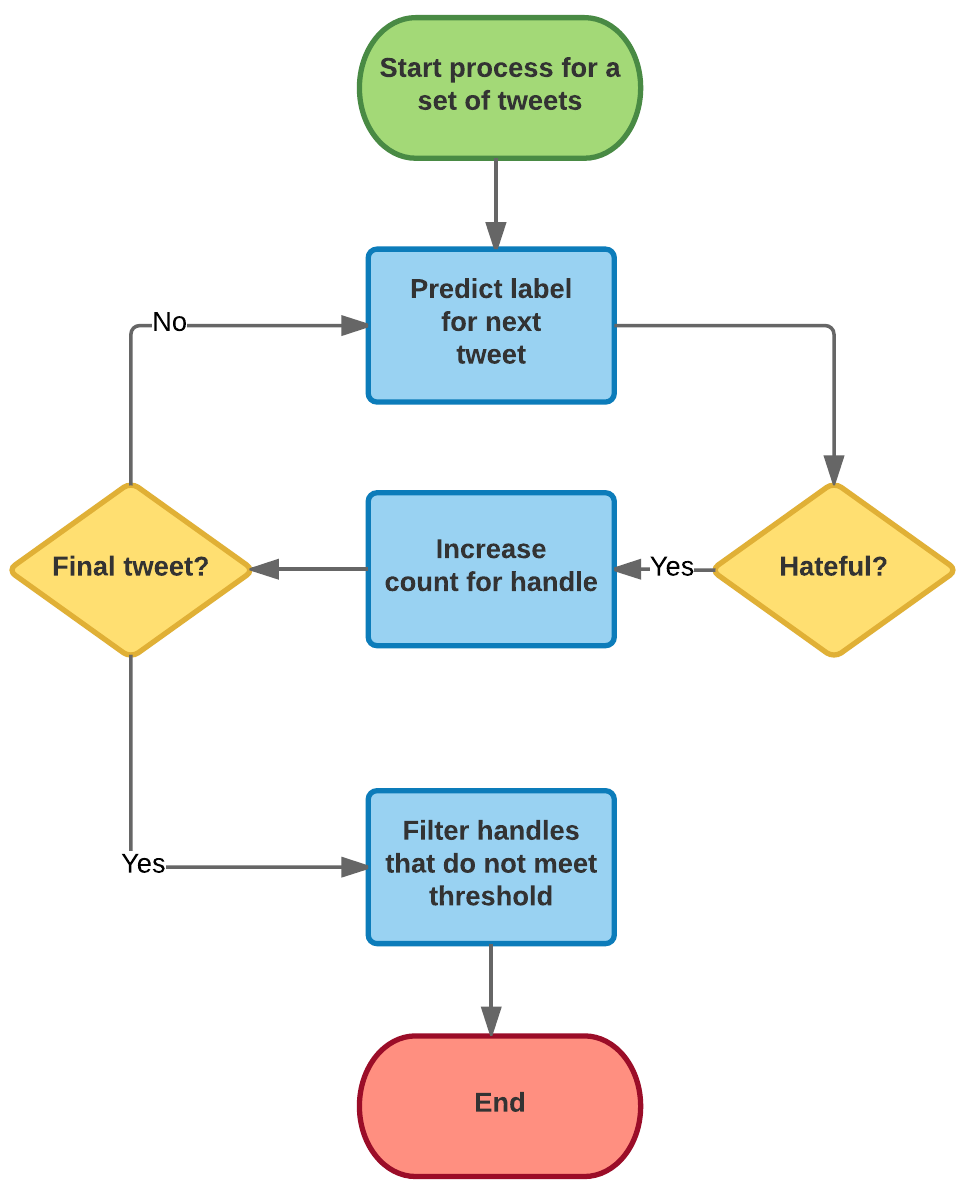}
\caption{Flowchart for identification process.}
\centering
\end{figure}

Since we wanted to train the classifier to be able to categorize a record of unstructured data i.e. an English sentence, we followed a bag of words model to represent a sentence in the training set with a boolean feature vector, where features are the most popular words in the corpora of training data. As part of the pre-processing of the tweets before they could be converted to feature vectors, we lemmatized every token in a tweet and then removed punctuation, stopwords, infrequent words from it. We built a support vector machine classifier to be able to segregate the hateful tweets. The next step involved running our classifier for a test set, which was created by randomly selecting 10\% of the posts in the tweet base. These sentences were then pre-processed and converted to feature vectors in the same manner as were the ones in the training set. We ran the classifier on this sample set to generate a list of tweets which were marked as hateful by the program. The final step involved finding the handles of all the tweets tagged as hateful and the number of hate tweets which had been posted from these handles. Handles having frequencies that crossed a certain threshold (decided upon based on heuristic) were considered qualified to be put in a list of potential aggressors.

\section{Experimental Results}
We used the annotated 1999 tweets, each containing single or multiple uses of code words, with 951 non-hateful tweets and 1048 hateful tweets as our training set. The annotated tweets were fed to the SVM classifier with a linear kernel and using 10 fold cross validation, to learn to distinguish between the two. We achieved an accuracy of 79.4397\% with a precision of 0.795 and recall of 0.794. More performance measures can be observed in Table 3. Furthermore, Table 4 showcases the confusion matrix obtained for the test.\\ 
 
\begin{table}[h!]
\centering
\caption{Classifier performance}
\hfill \break
\vspace{-5pt}
\begin{tabular}{|c| c| c| c| c|} 
\hline
Class & TP-Rate & FP-Rate & Precision & Recall \\ [0.5ex] 
\hline\hline
Benign & 0.752 & 0.167 & 0.803 & 0.752 \\
\hline
Hateful & 0.833 & 0.248 & 0.787 & 0.833 \\
\hline
Average & 0.794 & 0.21 & 0.795 & 0.794 \\
\hline
\end{tabular}

\label{table:3}
\end{table}

\begin{table}[h!]
\centering
\caption{Confusion Matrix}
\hfill \break
\vspace{-5pt}
\begin{tabular}{|c |c |c|} 
\hline
- & Benign & Hateful\\ [0.5ex] 
\hline\hline
Benign & 715 & 236 \\
\hline
Hateful & 175 & 873 \\
\hline
\end{tabular}

\label{table:4}
\end{table}

We ran the classifier on 23,401 tweets (which were the english language tweets that had further been extracted from 50,000 randomly selected tweets from our tweet base), resulting in the generation of handles that the classifier identified as having posted tweets that used the racist code, along with the count of the total number of corresponding hateful tweets that were associated with the user.

We found that for the given sample size, a threshold frequency of 4 was optimal in capturing a significant number of aggressors with as few false positives as possible.

\section{Conclusions}
We were successfully able to project the hate content problem on Twitter into a classification problem, that we solved to an extent using SVM. Additionally, apart from the system’s ability to predict for a given tweet whether a it is hateful or not, the system also generates a list of users who frequently post such content. This provides us with an interesting insight into the usage pattern of hate-mongers in terms of how they express bigotry, racism and propaganda.

A few different paths of possible research emerge from this project. First and foremost is the line of study we plan to undertake in the immediate future with regards to this project, which is to create a process to automatically identify new code words. This would require a large time frame because usage patterns change over long periods of duration, which imply a slow, gradual adoption of new code words in the lexicon. A study conducted after the generation of the list of aggressors indicated no use of new terminology. 

Another possible direction is to identify whether there exists some kind of network structure between the users i.e. if the users are connected, follow the same personalities etc. The hypothesis is there indeed exists some kind of nexus, although this needs to be verified. Community detection in this manner would provide an extra boost to identifying more aggressors and thereby their usage patterns.

We noted how there was a sudden upsurge in the use of code words during the second debate. More work can be carried out to determine if rate of generation of such coded messages peaks during turmoil, such as political unrest or terror attacks. As an extension to the above, research can also be done to check if there are certain regions with abnormal, consistently high uses of the code lexicon.

\nocite{*}
\bibliographystyle{aaai}
\bibliography{Magu_Joshi_Luo.bib}

\end{document}